\RequirePackage{fix-cm}
\documentclass{svjour3}                     
\smartqed  
\usepackage{graphicx}
\usepackage{cite}
\usepackage{array,multirow,makecell}
\usepackage{amsmath,amssymb,fancybox}
\usepackage[ruled,vlined,linesnumbered]{algorithm2e}
\usepackage{lineno,hyperref}
\usepackage[bottom]{footmisc}
\usepackage{float}
\usepackage{algorithmic}
\begin{document}

\title{Immunization of networks with non-overlapping community structure}


\author{Zakariya Ghalmane  \and Mohammed El Hassouni \and Hocine Cherifi}


\institute{Zakariya Ghalmane \at
              LRIT URAC No 29, Faculty of Science, Rabat IT center, Mohammed V University, Rabat, Morocco\\
              \email{zakaria.ghalmane@gmail.com}           
           \and
           Mohammed El Hassouni \at
              LRIT URAC No 29, Faculty of Science, Rabat IT center, Mohammed V University, Rabat, Morocco\\
              \email{mohamed.elhassouni@gmail.com}           
              \and
              Hocine Cherifi \at
              LE2I UMR 6306 CNRS, University of Burgundy, Dijon, France\\
              \email{hocine.cherifi@u-bourgogne.fr}           
}

\date{Received: date / Accepted: date}

\maketitle

\begin{abstract}
Although community structure is ubiquitous in complex networks, few works exploit this topological property to control epidemics. In this work, devoted to networks with non-overlapping community structure (i.e, a node belongs to a single community), we propose and investigate three deterministic immunization strategies.  In order to characterize the influence of a node, various pieces of information are used such as the number of communities that the node can reach in one hop, the nature of the links (intra community links, inter community links), the size of the communities, and the interconnection density between communities. Numerical simulations with the Susceptible-Infected-Removed (SIR) epidemiological model are conducted on both real-world and synthetic networks. Experimental results show that the proposed strategies are more effective than classical deterministic alternatives that are agnostic of the community structure. Additionally, they outperform stochastic and deterministic strategies designed for modular networks.
\keywords{Community structure \and Immunization strategy \and Epidemic spreading \and Influence \and Centrality \and SIR model}
\end{abstract}
\BlankLine
\BlankLine
\section{Introduction}

The structure of networks is crucial in explaining epidemiological patterns.  In the past few years, many immunization strategies have been developed using various topological properties of the network in order to mitigate and control the epidemic outbreaks. Despite the fact that there is clear evidence that many social networks show marked patterns of strong community structure  \cite{com1,com2,jebabli1,jebabli2}, this property has generally been ignored. A network with a strong community structure consists on cohesive subgroups of vertices that share many connections with members of their group and few connections with vertices outside their group. Bridge nodes are the ones that link different communities. They create a pathway of spreading disease outside of their community. Their  influence on epidemic spreading has been particularly investigated in previous works \cite{com4,gupta1,chakraborty}. Indeed, immunization of these nodes allows confining the disease into the community where it starts. However, one must not neglect the importance of the highly connected nodes embedded into  their community on the epidemic spreading.

Motivated by the reasons referred to above, we propose to make a better use of the information about the community structure in order to develop new immunization strategies. In this work, we restrict our attention to networks where each node belongs to a single community. We use a global deterministic approach. In other words, for each node of the network, an influence measure is computed and the nodes are ranked and immunized according to this measure. The first proposed method targets nodes having a big inter-community influence. It is measured by the number of neighboring communities linked to the node. The second immunization method targets nodes which could have at the same time a high influence inside and outside their communities. Greater importance is given to those belonging to large communities since they could affect more nodes. This strategy is based on a weighted combination of the number of intra-community and inter-community links of each node in the network. The third method has the same objectives than the previous one. It is designed in order to take also into account the density variation of the communities.

Experiments on real-world networks as well as synthetic networks have been conducted in order to evaluate the efficiency of the methods as compared to both deterministic and stochastic alternatives. Results of the investigations on the SIR epidemiological model show that the proposed strategies are in most cases, more efficient than their competitors. Extensive tests show that none of the proposed strategies outperform the two others in every situations. Indeed, controlled variations of the community structure in synthetic networks show that performances are greatly influenced by the strength of the community structure. The proposed strategies are thus clearly complementary. Indeed, their efficiency is at the best for networks belonging to a different range of community structure strength.

The remainder of this paper is organized as follows. In Section 2,  related works and immunization strategies are introduced. In Section 3, the proposed community-based strategies are defined. Section 4 introduces the experimental setting used in this work. In Section 5, the experimental results are presented. Finally, section 6 serves as a conclusion to the paper.
\section{Related work}
 
 Immunization strategies aim to immunize a few key nodes to achieve effectively the goal of reducing or stopping the spread of infectious diseases. They can be classified as stochastic or deterministic strategies. In stochastic strategies, local information about randomly selected nodes is used in order to identify target nodes. As they do not need any information about the full network structure, they can be used in situations where it is unavailable.  In deterministic strategies,  for each node of the network, one compute a measure of influence using local or global information. Nodes are then ranked and immunized according to their influence value.  Recently, researchers have begun to pay more attention to the community structure in terms of epidemic dynamics \cite{palla,fortunato}. Stochastic and deterministic strategies based on the community structure characteristics have been proposed. They can be categorized into two groups. The first group is based on topological properties of non-overlapping community structure, while the second group uses the overlapping community structure features  (i.e, a node could belong to multiple communities). We give a brief overview covering both types of strategies in order to highlight how communities can be advantageously used. However, experimental investigations are restricted to strategies designed for non-overlapping community structure which represents the focus of this study.

\subsection{Stochastic immunization strategies.}
These strategies target the most influential nodes using local information around randomly selected nodes. Their main advantage is that they require only a limited amount of information about the network topology. We present two stochastic methods based on  non overlapping community structure and one strategy designed for overlapping communities.\\

\noindent\textbf{- Community Bridge Finder (CBF):}  Proposed by Salathe et al. \cite{cbf}, it is a random walk based algorithm designed to search for bridge nodes. The basic idea is that real-world networks exhibit a strong community structure with few links between the communities. Therefore, when a walker reaches a node in another community, he is no longer linked to previously visited sites.\\
The CBF algorithm works as follows: 

Step 1: Select a random node $v_{i = 0}$ and follow a random path.

Step 2: $v_{i-1_{(i \succeq 2)}}$ is considered as a potential target if there is not more than one connection from $v_{i}$ to any of the previous visited nodes.

Step 3: Two random neighboring nodes of $v_{i}$ are picked (other than $v_{i-1}$), if there is no connections back to the previously visited nodes $v_{j \prec i}$ then, the potential target is marked as a bridge and it is immunized. Otherwise, a random walk at $v_{i-1}$ is taken back.\\ 
Comparisons have been performed with the Acquaintance strategy (A node is selected at random and one of its randomly selected neighbors is immunized). Extensive tests conducted on real-world and synthetic networks using the SIR epidemic model show that CBF performs mostly better, often equally well, and rarely worse than the Acquaintance strategy \cite{acq}. It performs particularly well on networks with strong community structure.\\
\textbf{- Bridge-Hub Detector (BHD):} The Bridge-Hub Detector \cite{bhd} is another variant of CBF strategy. It targets bridge hub nodes for immunization by exploring friendship circles of visited nodes. %
The procedure of the BHD algorithm can be specified as follows:

Step 1: Select a random node $v_{i = 0}$ and follow a random path.

Step 2: Let $v_{i}$ be the node selected after $i$ walks, and $f_{i}$ be the set of all neighbors of the node $v_{i}$. The node $v_{i}$ is targeted for immunization if there is at least a node in $f_{i}$ that is not a member in the set $F_{i-1}$ and that is not connected to the nodes in $F_{i-1}$ where $F_{i-1}\equiv f_{0} \bigcup f_{1} \bigcup f_{2} \bigcup. . .\bigcup f_{t-1}$. Otherwise, $v_{i}$ will not be targeted for immunization and $F_{i}$ will be updated to $F_{i}\equiv F_{i-1} \bigcup f_{i}$.

Step 3: One node $v_{H}$ is randomly selected for immunization among the nodes in $f_{i}$ that do not belong and could not be linked back to $F_{i-1}$.\\
Therefore, a pair of nodes, a bridge node and a bridge hub, are targeted for immunization via a random walk. BHD was applied on simulated and empirical data constructed from social network of
five US universities. Experimental results demonstrate that it compares favorably with Acquaintance and CBF strategies. Indeed, it results in reduced epidemic size, lower peak prevalence and fewer nodes need to be visited before finding the target nodes.\\ %
\textbf{- Random-Walk Overlap Selection (\textit{RWOS}):} This random walk based strategy \cite{rwos} targets the high degree overlapping nodes. It starts with defining the list of overlapping nodes obtained from known or extracted communities. Then, a random walk is followed starting from a random node of the network. At each step, the visited node is nominated as a target for immunization if it belongs to the list of overlapping nodes, otherwise, the random-walk proceeds. 
Simulation results on synthetic and real-wold networks with the SIR epidemic model show that the proposed method outperforms CBF and BHD strategies. In some cases it has a smaller epidemic size compared to the membership strategy where overlapping nodes are ranked according to the number of communities they belong to. In particular, its  performance improves in networks with strong community structures and with greater overlap membership values. 

To summarize, results show that stochastic methods designed for networks with community structure are more efficient that classical stochastic strategies. One of the very important contribution of these works is to demonstrate that  it is important to better take into account the modular organization of real-world networks in order to develop efficient  immunization strategies. Note, however, that stochastic methods are not as efficient as deterministic ones. Their main advantage is that they do not require a full knowledge of the global structure of the network. 

\subsection{Deterministic immunization strategies}
Nodes are immunized according to a rank computed using a specific influence (centrality) measure. Nodes with high centrality are targeted for immunization as they are more likely to infect many other nodes. The majority of known methods make use of the structural information either at the microscopic or at the macroscopic level to characterize the node importance. These strategies such as Degree and Betweenness measures   are very effective but they require the knowledge of the topology of the entire network. Refer to \cite{centralities} for a comprehensive survey on the subject. Based on the fact that the influence of a node is closely linked to the network topological structure,  and that a vast majority of real-world networks exhibit a modular organization, some deterministic methods have been developed lately for such networks.\\

\noindent\textbf{- \textit{Comm} strategy:} N. Gupta et al \cite{comm} proposed a strategy called the \textit{Comm} measure of a node. It combines the number of  its intra-community links (links with nodes inside its community) and the number of its inter-community links (links with nodes outside its community) to rank nodes that are both hubs in their community and bridges between communities. In this measure, the number of inter-community links is raised up to power two while the number of the intra-community links is not raised to give more importance to bridges. Results on synthetic and real-world networks show that the \textit{Comm} based strategy can be more effective as degree and betweenness strategies. However, it gives significant importance to the bridges compared to the community hubs. Yet, the hubs are commonly believed to be also influential nodes as they can infect their many neighbors. In some cases, they may play a very major role in the epidemic spreading.\\
\textbf{- Membership strategy:} L. Hébert-Dufresne et al. \cite{membership} proposed an immunization strategy based on the overlapping community structure of networks. Nodes are targeted according to their membership number, which indicates the number of communities to which they belong. Experiments with real-world networks of diverse nature (social, technological, communication networks, etc.) and two epidemiological models show that this strategy is more efficient as compared to degree, coreness and betweenness strategies. Furthermore, its best performances are obtained for high infection rates and dense communities.\\
\textbf{- OverlapNeighborhood strategy (\textit{ON}):} M. Kumar et al \cite{manish} proposed a strategy based on overlapping nodes. It targets immediate neighbors of overlapping nodes for immunization. This strategy is based on the idea that high degree nodes are neighbors of overlapping nodes. Using a limited amount of information at the community structure level (the overlapping nodes), this strategy allows to immunize high degree nodes in their respective communities. Experiments conducted on four real-world networks show that this immunization method is more efficient than stochastic methods such as CBF \cite{bet1,bet2}, BHD and RWOS methods. It also performs almost as well as degree and betweenness strategies while using less information about the overall network structure.

Globally, experimental results demonstrate that the cited deterministic strategies allows to reach the efficiency of classical strategies  that are agnostic about the community structure with less information. However, there is still room for improvement. 

\section{Proposed measures}
In order to quantify the influence of a node in the diffusion process on community structured networks, we propose three measures that integrate various levels of information.

\subsection{Number of Neighboring Communities Measure}
The main idea of this measure is to rank nodes according to the number of communities they reach directly (through one link). The reason for targeting these nodes is that they are more likely to contribute to the epidemic outbreak towards multiple communities. Note that all the nodes that do not have inter-community links share the same null value for this measure. 

\subsubsection{Definition:}
For a given node \textit{i} belonging to a community $C_{k}\subset C$,  the Number of Neighboring Communities $\beta_{c_{1}}(i)$ is given by:
\begin{equation}
 \beta_{c_{1}}(i) = \displaystyle\sum\limits_{C_{l}\subset C\backslash\{ C_{k} \}}  \  \bigvee_{j \in C_{l}} a_{ij}
\end{equation}\\
Where $a_{ij}$ is equal to 1 when a link between nodes $i$ and $j$ exists, and zero otherwise.\\
$\bigvee$ represents the logical operator of disjunction, i.e, $\bigvee_{j \in C_{l}} a_{ij}$ is equal to 1 when the node $i$ is connected to at least one of the nodes $j \in C_{l}$.

\subsubsection{Algorithm:}

\begin{algorithm}[H]
\DontPrintSemicolon
\SetAlgoLined
\SetKwInOut{Input}{Input}\SetKwInOut{Output}{Output}
\Input{Graph \textit{G(V,E)}, Community set: $C=\{C_{1},...,C_{m}\}$ where $m$ is the number of communities}
\Output{A map $M(node, measure \; value)$}

\BlankLine
Create and initialize an empty map $M(node, value)$\\
\For{each community $C_{k, k\in\{1,2,...,m\}}$ in the set \textit{C}}{ 
\For{each \textit{i} in the community $C_{k}$}{    
    
    \BlankLine
   
   $\beta_{C1} \leftarrow  0$
    \BlankLine
  \For{each \textit{C'} in \textit{$C \setminus \{C_{k}\}$}}{    
 
    \BlankLine
    \If{ $\exists j \in C': e_{ij}=1$ (a link exist between i and j)}{
    	\BlankLine   
  		 $\beta_{C1} \leftarrow  \beta_{C1}+1$
   } 
    }
    \BlankLine
    $ M.add(i,\beta_{C1})$
}}
\caption{Computation of the Number of Neighboring Community Measure}
\end{algorithm}

\subsubsection{Toy example:}

\begin{figure*}[t!]
\begin{center}
\includegraphics[width=13cm,height=3.2cm]{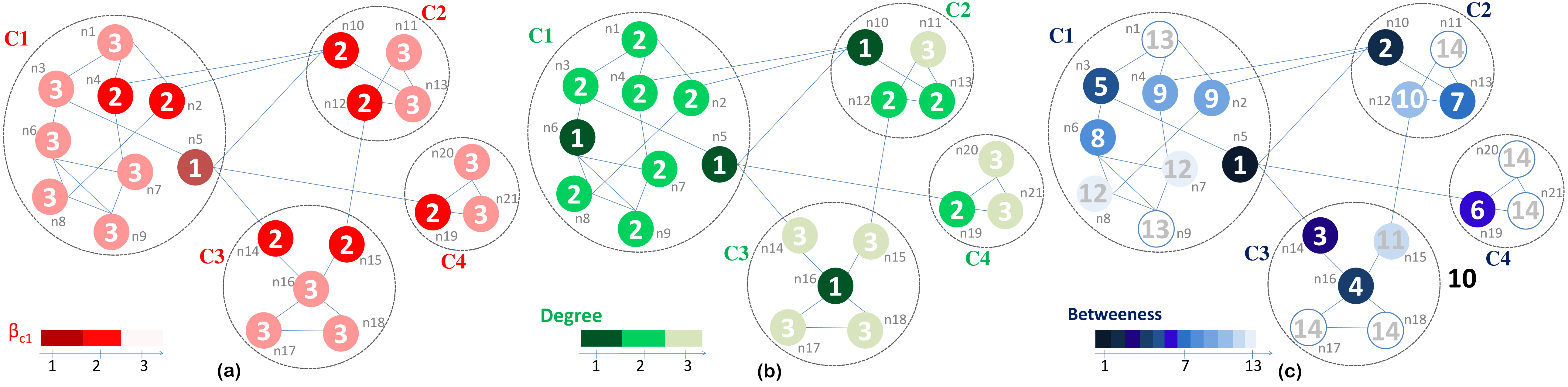}
\caption{ Rank of nodes according to (a) Number of Neighboring Communities measure $\beta_{c_{1}}$ (b) Degree measure (c) Betweenness measure. Nodes are ranked from the most influential (nodes having the highest measure value) to the less influential node (nodes having the lowest measure value) in the network.
\label{beta1}
}
\end{center}
\end{figure*}

In order to illustrate the behavior of this measure a toy example is given in \autoref{beta1}. Nodes are ranked according to the Number of Neighboring Communities measure  in \autoref{beta1} (a). This measure could distinguish between the different community bridges within the network where it targets the most influential ones for immunization. To clarify this idea, let's take the example of nodes \textit{n5} and \textit{n10} which are both community bridges and which have the same number of either internal and external links. According to Degree centrality measure in \autoref{beta1} (b), both nodes have the same rank since it depends only on their number of neighbors. However, they have different ranks according to the Number of Neighboring Communities measure. The proposed measure gives  more importance to node \textit{n5} which is linked to three external communities, so ever if it is contaminated, it can transmit the epidemic disease first  to its own community \textit{C1} and also towards the neighboring communities \textit{C2}, \textit{C3} and \textit{C4}. While the epidemic disease could be transmitted to nodes belonging to the communities \textit{C1} and \textit{C2} in the case of node \textit{n10} contamination. Therefore, \textit{n5} represents a threat for 20 nodes which is the majority of nodes of the network, while \textit{n10} represents a threat to only 12 nodes. On the other hand, Betweenness centrality measure tries to immunize network bridges having a high fraction of geodesics passing through them. Yet, these nodes do not always represent the community bridges which are very important and must be immunized in priority in this type of network.  As it is shown in \autoref{beta1} (c) according to Betweenness measure, the nodes \textit{n15} and \textit{n12} are ranked among the less influential nodes, although, both are community bridges that are likely to contribute to the epidemic outbreak to external communities. Therefore, the Number of Neighboring Communities measure targets the most influential bridges which can spread the epidemics to multiple communities. 

\subsection{Community Hub-Bridge Measure}
Each node of the network share its links with nodes inside its community (intra-community links) and nodes outside its community (inter-community links). Depending of the distribution of these links, it can propagates the epidemic more or less in its community or to its neighboring communities. Therefore, it can be considered as a hub in  its community and a bridge with its neighboring communities. That is the reason why we call this measure the Community Hub-Bridge measure. Furthermore, the hub influence depends on the size of the community, while the bridge influence depends on the number of its neighboring communities.

\subsubsection{Definition:}

For a given node \textit{i} belonging to a community $C_{k}\subset C$ , the Community Hub-Bridge measure $\beta_{c_{2}}(i)$ is given by:
\begin{equation}
\beta_{c_{2}}(i)_{i\in C_{k}}= h_{i}(C_{k}) +  b_{i}(C_{k})
\end{equation}
Where:
\begin{equation} 
h_{i}(C_{k})= Card(C_{k}) * k_{i}^{intra}(C_{k})
\end{equation} 
\begin{equation}
b_{i}(C_{k})= \beta_{c1}(i) * k_{i}^{inter}(C_{k})
\end{equation}

$k_{i}^{intra}(C_{k})$ and $k_{i}^{inter}(C_{k})$ are respectively the intra-community degree and the inter-community degree of node \textit{i}. $Card(C_{k})$ is the size of its community.  $\beta_{c1}(i)$ represents the number of its neighboring communities.\\ 
$h_{i}(C_{k})$ tend to immunize preferentially hubs inside large communities. Indeed, they can infect more nodes than those belonging to small communities.\\
 $b_{i}(C_{k})$ allows to target nodes that have more links with various communities. Such nodes have a big inter-community influence. 

\subsubsection{Algorithm:}

\begin{algorithm}[H]
\DontPrintSemicolon
\SetAlgoLined
\SetKwInOut{Input}{Input}\SetKwInOut{Output}{Output}
\Input{Graph \textit{G(V,E)}, Community set: $C=\{C_{1},...,C_{m}\}$ where $m$ is the number of communities}
\Output{A map $M(node, measure \; value)$}

\BlankLine
Create and initialize an empty map $M(node, value)$\\   

\For{each community $C_{k, k\in\{1,2,...,m\}}$ in the set \textit{C}}{
Calculate the size of the community $C_{k}$\\ 
\For{each \textit{i} in the community $C_{k}$}{   
    \BlankLine
Calculate $\beta_{C1}$ the number of neighboring communities of $i$ described in Algorithm 1\\
Calculate $k_{i}^{intra}(C_{k})$ the intra-community links of the node \textit{i}\\
Calculate $k_{i}^{inter}(C_{k})$ the inter-community links of the node \textit{i}
	\BlankLine
    \BlankLine
Calculate $\beta_{C2}$ the Community Hub-Bridge measure of the node $i$ according to Equation (2)     
    \BlankLine
    \BlankLine
    $ M.add(i,\beta_{C2})$
}}
\caption{Computation of the Community Hub-Bridge measure}
\end{algorithm}

\subsubsection{Toy example:}

\begin{figure*}[t!]
\begin{center}
\includegraphics[width=13cm,height=3.2cm]{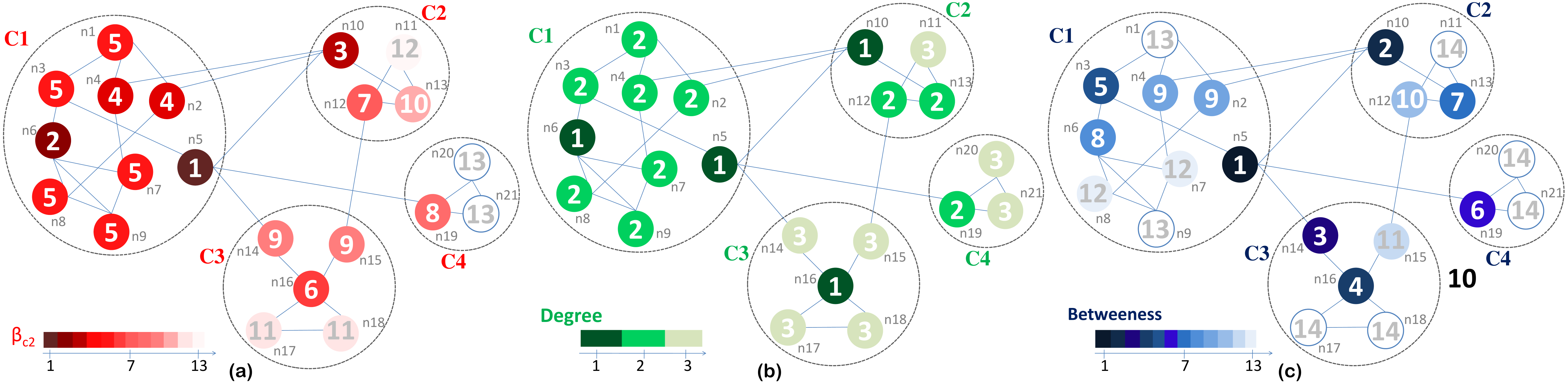}
\caption{ Rank of nodes according to (a) Community Hub-Bridge measure $\beta_{c_{2}}$ (b) Degree measure (c) Betweenness measure. Nodes are ranked from the most influential to the less influential node in the network.
\label{beta2}
}
\end{center}
\end{figure*}

\autoref{beta2} (a) shows the rank of nodes according to the Community Hub-Bridge measure. This measure targets nodes having a good balance of inner and outer connections inside the community. However, it takes into consideration the size of the community for quantifying the influence of the community hubs and the number of neighboring communities for the bridges. Thus, even-though, both \textit{n6} and \textit{n16} have four inner links inside their own communities \textit{n6} is considered more influential because it is located in community \textit{C1} which is the largest community of the network. Therefore, it could be a threat to several nodes inside the network if ever it is infected. Unlike degree measure in \autoref{beta2} (b) that classifies the nodes \textit{n6} and \textit{n16} in the same rank based on their number of connections without considering their location within the network.
It is also noticed from the previous example given in \autoref{beta1} (a) that many nodes have the same rank because they have the same number of neighboring communities. So, if we consider the nodes \textit{n10} and \textit{n12}, they are both connected to only one neighboring community (respectively \textit{C1} and \textit{C3}), consequently they have the same rank. However, \textit{n10} has a bigger connectivity to \textit{C1} in term of the number of outer links. The reason why we introduced the quantity of outer links as a new parameter in the second term of the Community Hub-Bridge measure to distinguish between bridges having big connectivity and those having lower connectivity with external communities. Based on Community Hub-Bridge measure \textit{n10} is more influential than node \textit{n12} as it can be seen in \autoref{beta2} (a) since it has three outer connections towards community \textit{C1} while node \textit{n12} has only one connection towards \textit{C3}. Therefore, the influence of nodes according to this measure is linked to two factors; the importance of nodes inside their communities by giving the priority to those located in large communities, and the connectivity of the nodes towards various communities.

\BlankLine
\BlankLine
\BlankLine
\BlankLine
\BlankLine
\BlankLine
\BlankLine
\BlankLine
\BlankLine

\subsection{Weighted Community Hub-Bridge Measure}

The Community Hub-Bridge measure targets in priority the hubs in large communities and the bridges linked to multiple communities. However, no importance is given to the community structure strength.  When the community structure is well-defined, more importance should be given to the bridges. Indeed, in this case breaking the network in multiple communities allows to contain the epidemic spreading where it started. On the contrary, when the community structure is very loose, it is of prime interest to immunize the hubs in large communities. Weighting each component  of the community Hub-Bridge allows therefore to give more or less importance to bridges or hubs according to the  community structure strength.

\subsubsection{Definition:}

For a given node \textit{i} belonging to a community $C_{k}\subset C$, the Weighted Community Hub-Bridge Measure $\beta_{c_{3}}(i)$ is given by:
\begin{equation}
\beta_{c_{3}}(i)_{i\in C_{k}}=\rho_{C_{k}} * h_{i}(C_{k}) + (1 - \rho_{C_{k}}) * b_{i}(C_{k})
\end{equation}\\
Where $\rho_{C_{k}}$ represents the interconnection density between the community $C_{k}$ and the other communities of the network. It is given by: 
 \begin{equation}
\rho_{C_{k}}=\dfrac{\sum\limits_{i \in C_{k}} k^{inter}_{i}/(k^{inter}_{i}+k^{intra}_{i})}{Card(C_{k})}
\end{equation} 
If the communities are very cohesive, then more importance is given to the bridges in order to isolate the communities. Otherwise, more importance is given to the hubs inside large communities. 
\subsubsection{Algorithm:}

\begin{algorithm}[H]
\DontPrintSemicolon
\SetAlgoLined
\SetKwInOut{Input}{Input}\SetKwInOut{Output}{Output}
\Input{Graph \textit{G(V,E)}, Community set: $C=\{C_{1},...,C_{m}\}$ where $m$ is the number of communities}
\Output{A map $M(node, measure \; value)$}

\BlankLine
Create and initialize an empty map $M(node, value)$\\   

\For{each community $C_{k, k\in\{1,2,...,m\}}$ in the set \textit{C}}{
Calculate the size of the community $C_{k}$\\
Calculate $\rho_{C_{k}}$ the the interconnection density between the community $C_{k}$ and the other communities of the network according to Equation (6)\\ 
\For{each \textit{i} in the community $C_{k}$}{   
    \BlankLine
Calculate $\beta_{C1}$ the number of neighboring communities of $i$ described in Algorithm 1\\
Calculate $k_{i}^{intra}(C_{k})$ the intra-community links of the node \textit{i}\\
Calculate $k_{i}^{inter}(C_{k})$ the inter-community links of the node \textit{i}
	\BlankLine
    \BlankLine
Calculate $\beta_{C3}$ the Weighted Community Hub-Bridge measure of the node $i$ according to Equation (5)     
    \BlankLine
    \BlankLine
    $ M.add(i,\beta_{C3})$
}}
\caption{Computation of the Weighted Community Hub-Bridge measure}
\end{algorithm}

 \subsubsection{Toy example:}
\begin{figure*}[t!]
\begin{center}
\includegraphics[width=13cm,height=3.2cm]{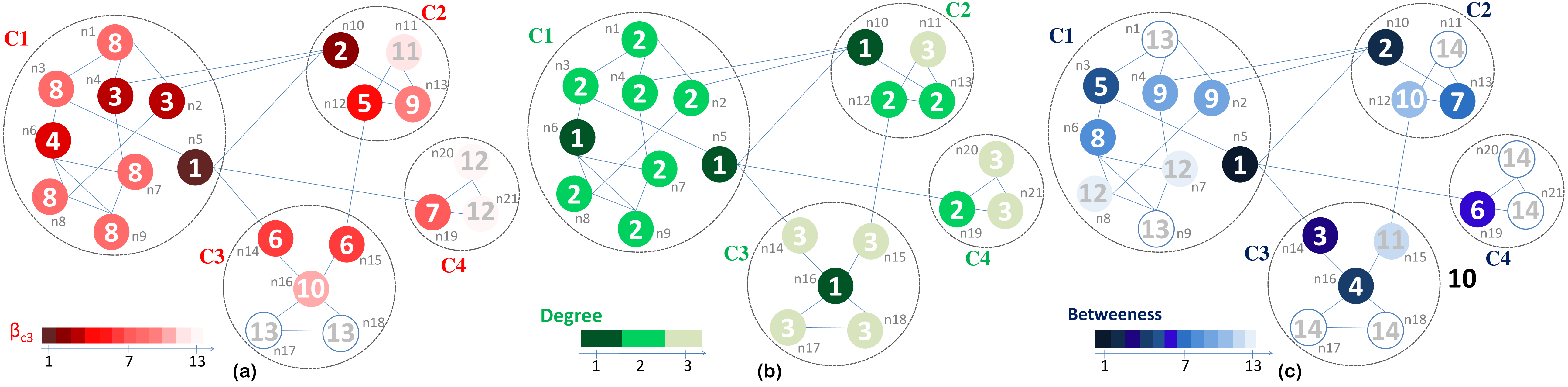}
\caption{ Rank of nodes according to (a) Weighted Community Hub-Bridge measure $\beta_{c_{3}}$ (b) Degree measure (c) Betweenness measure. Nodes are ranked from the most influential to the less influential node in the network.
\label{beta3}
}
\end{center}
\end{figure*}

This measure is very similar to the Community Hub-Bridge measure. However, it is designed in order to be able to adapt to the community 
structure strength. As it can be noticed from \autoref{beta3} (a) the network given in this example has a well-defined community structure. Thus, the interconnection density $\rho_{C_{k}}$ has a small value for all the communities of the network. As we can clearly see, if we take the example of the community \textit{C1}, the density of inter-community links is equal to $\rho_{C_{1}} \approx 0.15$. Consequently, 15\% of importance is given to the hub term $h_{i}(C_{1})$ and 85\% of importance is given to the bridge term $b_{i}(C_{1})$. This explains why all the community bridges (\textit{n5}, \textit{n2} and \textit{n4}) are immunized before the other nodes of the community \textit{C1}.  It helps to isolate this community and prevent the epidemic diffusion to move from \textit{C1} to the other communities of the network. Thus,  the Weighted Community Hub-Bridge measure has the ability to adapt to the strenght of the community structure. It gives more weight to the bridges when the network has a well-defined community structure in order to isolate the communities, while it gives more weight to hubs in the case of networks with a weak community structure since the network acts in this case like a single big community.

\section{Experimental Setting}
In this section, we present the data and methods used in the empirical evaluation of the various immunization strategies presented above.

\subsection{Datasets}

In order to evaluate the various measures under study, synthetic networks with controlled topological properties, together with  real-world networks have been used.
 \subsubsection{ Synthetic networks}

 Synthetic networks are generated using the LFR (Lancichinetti, Fortunato and Radicchi) algorithm \cite{lfr}. It generates random samples of networks with power-law distributed degree and community size. Hence, LFR algorithm guarantees networks with realistic features \cite{orman3}. This algorithm allows to control different parameters when generating networks. Mainly, the mixing parameter $\mu$, determines the ratio of the number of external neighbors of a node to the total degree of the node. Its value controls the strength of the community structure. For small values of $\mu$, the communities are well-separated because they share  few links, whereas when $\mu$  increases the proportion of inter community links becomes higher, making community identification a difficult task.
 Experimental studies showed that for a scale-free network, the degree distribution exponent $\alpha$  usually ranges from 2 to 3, and the maximal degree is estimated to be $k_{max} \sim n^{1/(\alpha-1)}$ \cite{barabasi,boccaletti,newman}. The parameters values used in our experiments are given in \autoref{t1}. 
 
 \begin{table}
   \centering
    \caption{LFR network parameters}
        \label{t1}	
      \begin{tabular}{lc}
      \hline       

	Number of nodes & 15 000 \\
	Average degree & 7 \\
	Maximum degree & 122 \\
	Exponent for the degree distribution & 3 \\
	Exponent for the community size distribution & 2.5 \\
	Mixing parameter & 0.1, 0.4, 0.7, 0.9 \\
	Community size range set & [50 250],[100 500] \\
	\hline
      \end{tabular}%
\end{table}
 \subsubsection{ Real-world networks}

Real-world networks of various nature (online social networks, a technological network and a collaboration network)  are used in order to test the immunization strategies.\\
\textbf{- Facebook:} We use a network gathered by Traud et al. \cite{traud} from Facebook  \footnote{\url{http://code.google.com/p/socialnetworksimulation/}} online social network . This data includes the friendship network of five universities in the US.  It provides also information about the individuals such as the dormitory, the major or the field of specialization and the year of class.\\
\textbf{- Power-grid:}  This technological network is an undirected, unweighted network containing information about the topology of the Western States Power Grid of the United States. An edge represents a power supply line. A node is either a generator, a transformer or  a substation. This data\footnote{\url{http://www-personal.umich.edu/~mejn/netdata/}} is compiled by D. Watts and S. Strogatz \cite{watts}. \\
\textbf{- General Relativity and Quantum Cosmology (GR-QC):} GR-QC\footnote{\url{http://snap.stanford.edu/data/ca-GrQc.html}} is a collaboration network collected from the e-print arXiv. It covers scientific collaborations between authors of papers submitted to the General Relativity and Quantum Cosmology category. The nodes represent the authors and there is a  link between two nodes if they co-authored a paper. This data is available in the SNAP repository compiled by J. Leskovec et al. \cite{leskovec}.\\
As the community structure of these networks is unknown, we use a community detection algorithm. We choose to use the Louvain algorithm that proved to be efficient in synthetic networks for a wide range of community strength \cite{orman2}. Furthermore, the topological properties of the uncovered communities are also realistic \cite{orman1}.

The basic topological properties of these networks are given in \autoref{t2}.

\begin{table}
   \centering
    \caption{The basic topological properties of six real-world networks. \textit{N} and \textit{E} are respectively the total numbers of nodes and links. \textit{Q} is the modularity. $N_{c}$ is the number of communities. }
        \label{t2}	
      \begin{tabular}{lcccc}
      \hline       

	Network & \textit{N} & \textit{E} & \textit{Q} & $N_{c}$ \\
	\hline
	Caltech & 620 & 7255 & 0.788 & 13\\
	Princeton & 5112 & 28684 & 0.753 & 21\\
	Georgetown & 7651 & 163225 & 0.662 & 42\\
	Oklahoma & 10386 & 88266 & 0.914 & 67\\
	Power grid & 4941 & 6594  & 0.93 & 41\\
	CR-QC & 5242 & 14496 & 0.86 & 396\\
	\hline
      \end{tabular}%
\end{table}

%

\subsection{Epidemiological model}
\begin{figure*}[t!]
\begin{center}
\includegraphics[width=7.5cm,height=2.7cm]{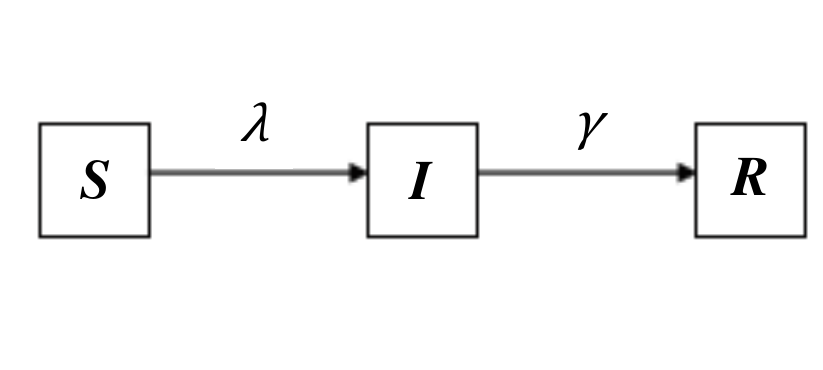}
\caption{The infection mechanism of the classic SIR model.}
\label{f2}
\end{center}
\end{figure*}

The Susceptible-Infected-Recovered (SIR) epidemic model \cite{sir1,sir2} is used to simulate the spreading process in networks. In this model, there are three states for each node: susceptible (S), infected (I) and recovered (R). The infection mechanism of the SIR model is shown in \autoref{f2}. Initially, targeted nodes are chosen according to a given immunization strategy until a desired immunization coverage of the population is achieved, and their state is set to resistant R. All remaining nodes are in S state. After this initial set-up, infection starts from a random susceptible node. Its state changes to I. At each time step, the epidemic spreads from one infected node to a neighboring susceptible node according to the transmission rate  of infection $\lambda$. Furthermore, infected nodes recover at rate $\gamma$, i.e. the probability of recovery of an infected node per time step  is $\gamma$. If recovery occurs, the state of the recovered node is set from infected to resistant. The epidemic spreading process ends when there is no infected node in the network. After each simulation, we record the total number of recovered nodes (the epidemic size). 
\\

\subsection{Immunization scheme}
\begin{figure*}[t!]
\begin{center}
\includegraphics[width=15cm,height=6cm]{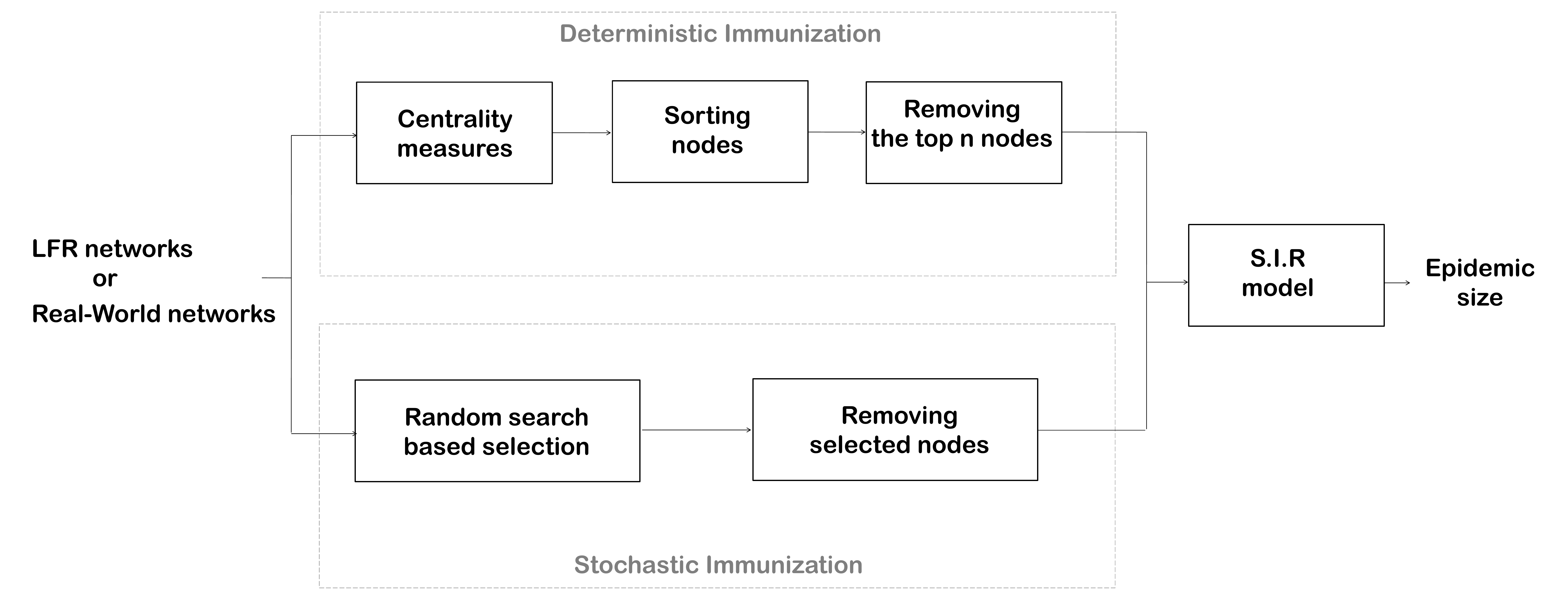}
\caption{The main steps of the immunization scheme.}
\label{f3}
\end{center}
\end{figure*}

To investigate the spread of an infectious disease on a contact network, we use the methodology described in \autoref{f3}. For deterministic strategies, the influence of every node in the network is calculated according to a given centrality measure. Then, nodes are sorted in decreasing order of their influence values. Next, nodes with highest centrality are removed from the network (or their state is set to resistant) until a desired immunization coverage is achieved. For stochastic immunization, nodes are targeted and removed according to a random strategy initiated from randomly chosen nodes in the network. In both cases, the network obtained after the targeted immunization is used to simulate the spreading process, running the SIR epidemic model simulations. After a simulation, we record the total number of cases infected (the epidemic size). In order to ensure the effectiveness of the SIR propagation model evaluation, results are averaged over 600 independent realizations. Finally, we calculate the mean epidemic size to evaluate the effectiveness of the proposed methods. 

\subsection{Evaluation Criteria}
To compare the performance of different immunization strategies, we use the fraction of the epidemic size. We also use the relative difference of outbreak size $\Delta r_{\beta_{0},\beta_{c}}$ defined by: 
\begin{equation}
\Delta r_{\beta_{0},\beta_{c}} = \dfrac{R_{\beta_{0}} \thinspace \_ \enspace  R_{\beta_{c}}}{R_{\beta_{0}}} 
\end{equation}
Where $R_{\beta_{0}}$ and $R_{\beta_{c}}$ are respectively the final numbers of recovered nodes for the alternative and the proposed strategy after the SIR simulations. If the relative difference of outbreak size is positive, the epidemic spreads less with the proposed strategy. Therefore, it is the most efficient one. Otherwise, the epidemic spreads more with the proposed strategy and the alternative strategy is more efficient. 
 
\section{Results and discussion} 

In this section, we report the results of two sets of experiments. The first set of experiments is performed with synthetic networks with controlled community structure. It is aimed at getting a better understanding of the relationship between the community structure and the centrality measures. These experiments are conducted on networks generated with the LFR algorithm. Indeed, this algorithm allows to control various topological properties of the community structure. We investigate the influence of the strength of the community structure. The community size range effect is also studied. Finally, the proposed immunization strategies are compared with both deterministic strategies (Degree, Betweenness and Comm strategies) and stochastic strategies (Community Bridge Finder \cite{cbf} and Bridge Hub Detector \cite{bhd} strategies).\\
The second set of experiments concerns real-world networks. Online Social networks, a technological network and a collaboration network are used. Recall that, as there is no ground-truth data for these networks, the community structure is uncovered using the Louvain Algorithm. Indeed, previous studies on synthetic networks have shown that it succeeds in identifying the communities for a large range of community structure strength \cite{bhd}. First, the proposed immunization strategies are compared and discussed, then their evaluation is performed against both stochastic and deterministic alternative strategies.

\subsection{Synthetic networks}

\subsubsection{Influence of the strength of the Community Structure}
\begin{figure*}[t!]
\begin{center}
\includegraphics[width=15cm,height=10cm]{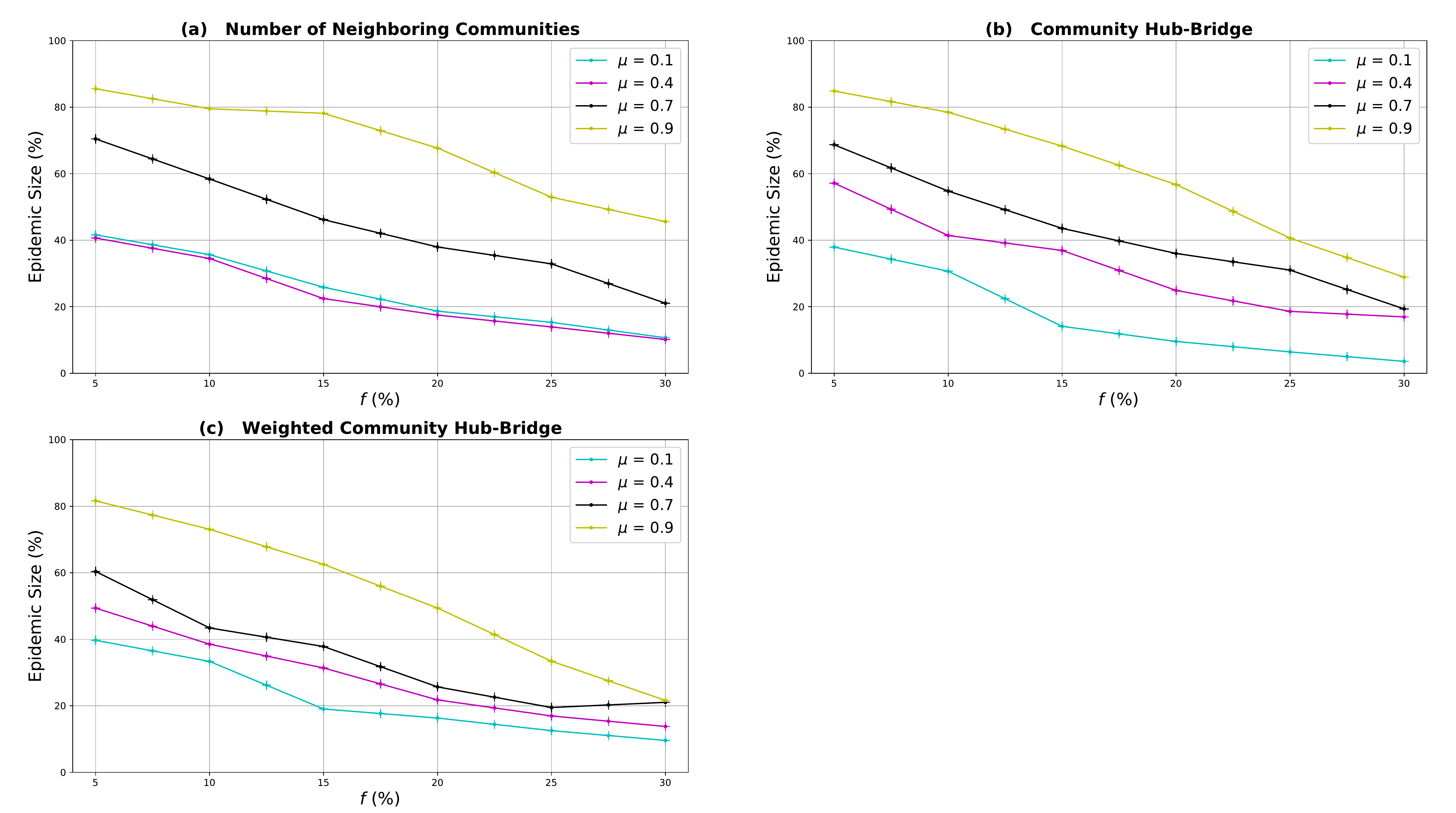}
\caption{Influence of the strength of the community structure on the epidemic size of the proposed methods. Simulations are performed on LFR-generated networks with various mixing parameter values $ \mu$. Each epidemic size value is the average of 600  S.I.R simulation runs.}
\label{r1}
\end{center}
\end{figure*}
In the LFR model, the mixing parameter value $ \mu$  varies from 0 to 1. It allows to control the strength of the community structure from well-separated communities with few inter-community links (low values of $ \mu$) to a network with no community structure (high values of $ \mu$).  In order to  investigate the effect of the strength of the community structure on the performance of the proposed methods, five networks have been generated for each value of the mixing parameter ($\mu=$ 0.1, 0.4, 0.7 and 0.9).  \autoref{r1} reports the average fraction of the epidemic size versus the proportion of immunized nodes for each $\mu$ value. It can be noticed from the results reported in this figure, that the performance of Community Hub-Bridge and Weighted Community Hub-Bridge exhibits a similar behavior.  Whatever the fraction of immunized nodes the methods, perform best when the communities are well-separated. When the fraction of inter-connections between the communities increases, performance decrease gradually. Indeed, with well-separated communities, the epidemics is localized to few communities, while it tends to spread more when the inter connections increase. The Number of Neighboring Communities strategy shows its best performance for a medium range community strength value ($\mu=0.4$). Its efficiency decreases slightly in the case of well-defined community structure, and it gets even worse when it is very loose. Let's now turn to the comparisons of the proposed methods between them. We can distinguish three cases depending of the community structure strength.

In networks exhibiting a very strong community structure, we can see in  \autoref{r1}  that the Community Hub-Bridge strategy  is the most efficient. This is due to the fact that both alternatives methods (Number of Neighboring Communities and  Weighted Community Hub-Bridge strategies)  target preferentially the bridges. In fact, it is not the best solution in a network where the intra-community links predominate.  As there is few external connections compared to the total connections (intra-community links are considered to be 90\% of the total links of the network when $\mu$=0.1), local outbreaks may die out before reaching other communities. Therefore, immunizing community hubs seems to be more efficient than immunizing bridges in networks with strong community structure. This is the reason why the Community Hub-Bridge method which targets nodes having a good balance of inner and outer connections is more efficient.

In networks with weak community structure as it can be seen in \autoref{r1}, the Weighted Community Hub-Bridge method is the most efficient. Indeed, when $\mu$ has a high value, the network does not have a well-defined community structure. In that case, Weighted Community Hub-Bridge strategy can better adapt to the community structure.  It gives more weight to the community hubs as they are the most influential nodes in networks with a loose community structure. Remember that the network acts as a single community in the extreme case where $\mu \geqslant 0.9$.  That is the reason why it performs better than the other proposed methods.

In networks with community structure of medium strength, the Number of Neighboring Communities outperforms all the other proposed methods as it reported in  \autoref{r1}. In this type of networks, nodes have many external connections while maintaining a well-preserved community structure. Therefore, there is much more options for the epidemic to spread easily to neighboring communities. As the Number of Neighboring Communities strategy targets the most influential community bridges, it prevents the epidemic spreading to multiple communities. This is the reason why this immunization method shows its best performance in this case.

To summarize,  Community-Hub bridge strategy is well-suited to situations where the communities are well-defined (Dense communities with few links between communities). One must prefer weighted Hub-Bridge strategy when the community structure is very loose. In between the Number of Neighboring Communities strategy is more efficient.
 
\subsubsection{Community Size Range effects}
\begin{figure*}[t!]
\begin{center}
\includegraphics[width=15cm,height=10cm]{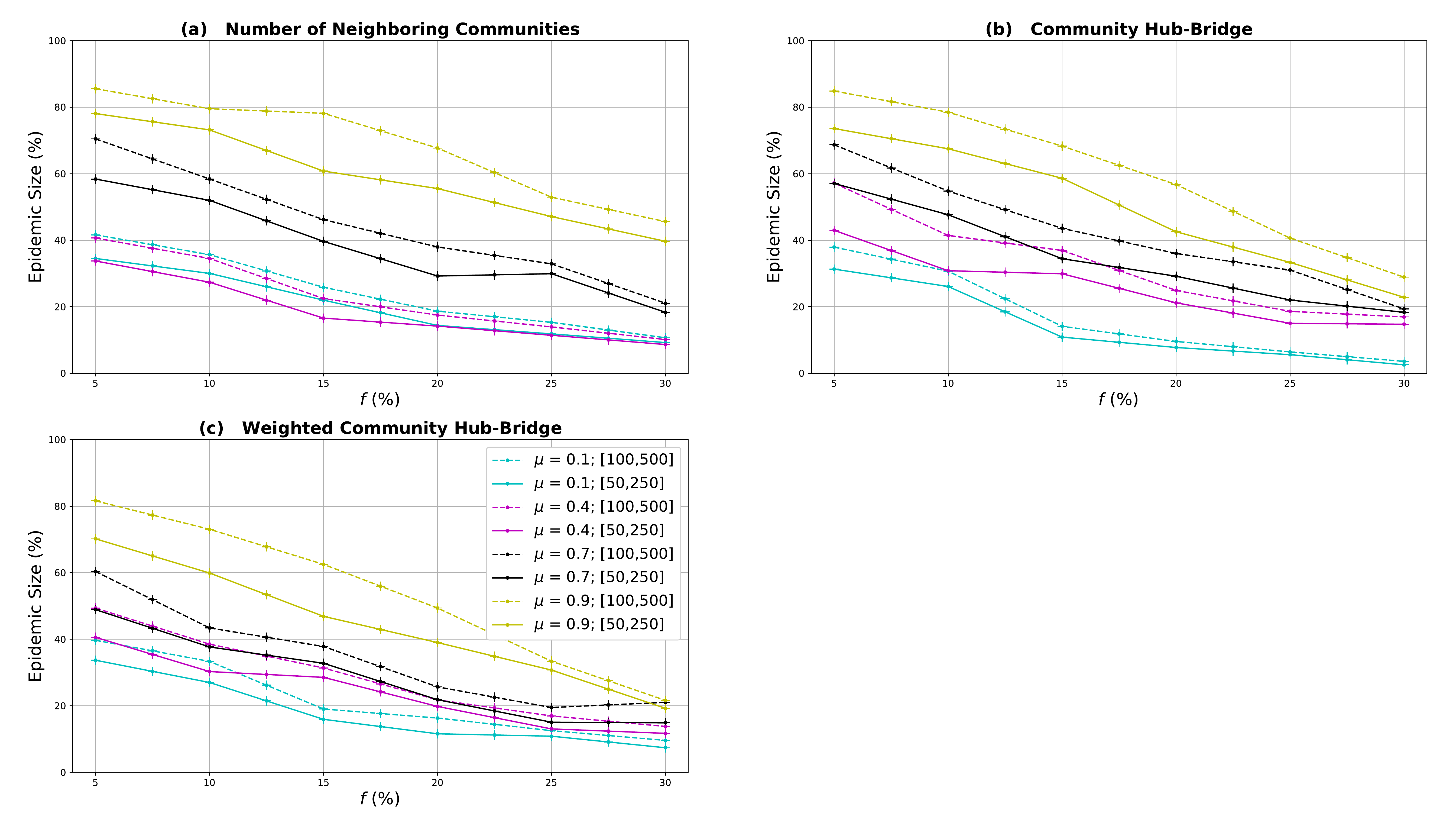}
\caption{Effect of community size range on the epidemic spreading of the proposed methods. Simulations are performed on LFR network with different community structure. Each value is the average of 600 runs per network and immunization method.}
\label{r2}
\end{center}
\end{figure*}

The aim of this investigation is to show the impact of the community size range on the performance of the proposed methods. Studies reported above have been performed  with community structure size in the range  [100, 500]. In this paragraph, they are also evaluated in networks with community size range equals to [50, 250].  \autoref{r2}  reports the epidemic size versus the percentage of immunized nodes for values of the mixing parameter $\mu$  ranging from $\mu=0.1$ to $\mu=0.9$,  and with the two community size range under study. One can see that all the immunization strategies exhibit the same behavior. They always perform better in networks with smaller community size range. Furthermore, the differences between the epidemic sizes in the two situations decrease when the proportion of immunized nodes increases. In networks with a big community size range, there are a small number of communities. Consequently,  the range of the Number of Neighboring Communities measure is also small, and many nodes have the same values (as it is shown in the example given in section 3.1). That makes the ranking less efficient. In networks with a smaller community size range, there are much more communities. In this case, more nodes have different numbers of neighboring communities values and the ranking is more efficient. That is the reason why the Number of Neighboring communities performs better in the latter case. Concerning Community Hub-Bridge and Weighted Community Hub-Bridge measures, both are weighted by the number of neighboring communities. This weight becomes more discriminative as the community size range decreases. That explains why they also perform better in networks with small community size range.

From \autoref{r2}, one can also see that the Community Hub-Bridge method is always the best immunization method in networks with well-defined community structure, and that the Number of Neighboring Communities outperforms the other proposed immunization methods (where $\mu=0.4$). Moreover, the Weighted Community Hub-Bridge is still the most efficient one in networks with non-cohesive community structure.  
 
\subsubsection{Comparison with the alternative methods}
\begin{figure*}[t!]
\begin{center}
\includegraphics[width=15cm,height=14cm]{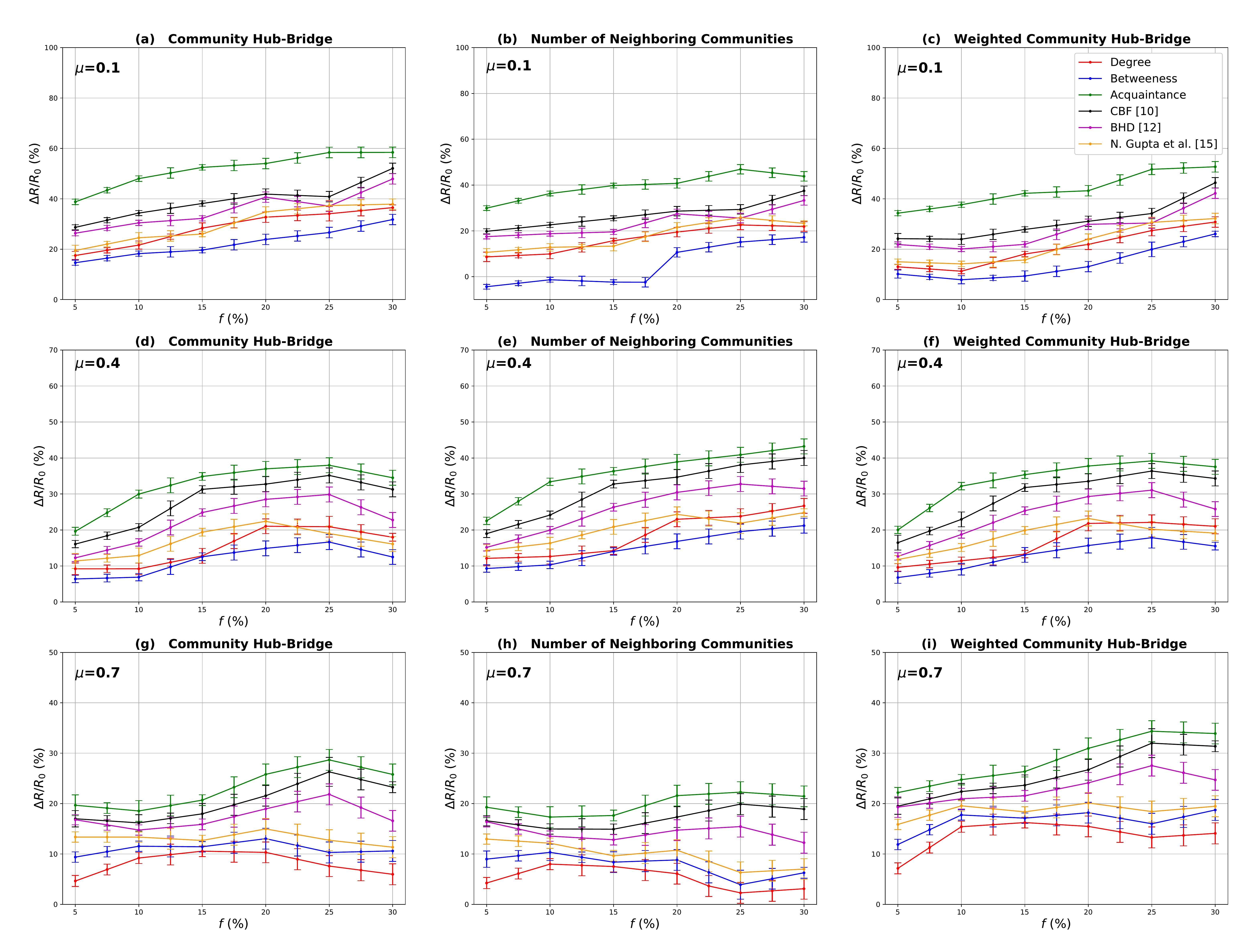}
\caption{The relative difference of the outbreak size $\Delta R/R_{0}$ as a function of the fraction of immunized nodes $f$. The left panels show the difference between Community Hub-Bridge method and the alternative methods. The middle panels show the difference between Number of Neighboring Communities method and the alternative methods, while the right panels show the difference between Weighted Community Hub-Bridge and the alternative methods. We note that a positive value of $\Delta R/R_{0}$ means a higher performance of the proposed method. Simulations are performed on LFR network with different community structure. Final values are obtained by running 600 independent simulations per network, immunization coverage and immunization method.}
\label{r3}
\end{center}
\end{figure*} 
\autoref{r3} reports the relative difference of the outbreak size between the proposed strategies and both stochastic (Acquaintance, CBF, BHD) and deterministic alternatives (Degree, Betweenness and Comm) as a function of the fraction of immunized nodes. 
Community Hub-Bridge is taken as the reference in (a), (d)  and (g), Number of Neighborhood communities in  (b), (e) and (h) and Weighted Community Hub-Bridge in (c), (f) and (e). The values of the mixing parameter ($ \mu=0.1$, $\mu=0.4 $, $\mu=0.7$) cover the three situations in terms of community strength (strong, median and weak community structure). 

\autoref{r3} (a), (d) and (g) shows that $\Delta R/R_{0}$ has always a positive value. Thus, Community Hub-Bridge always yields a smaller epidemic size compared to all the alternative methods whatever the fraction of immunized nodes values, and this holds for all the range of community structure strength. The middle panels of \autoref{r3} reports the results of the comparative evaluation of the Number of Neighboring Communities strategy. Overall, it is more efficient than the tested alternative methods. However, Betweenness performs better in networks with strong community structure ($\mu = 0.1$) and fraction of immunized nodes below 18\%. Indeed, the relative difference is negative in this case. Therefore, targeting the community bridges is not the best immunization solution in networks with very well-defined community structure. It can be also noticed from \autoref{r3} (c), (f) and (i) that the Weighted Community Hub-Bridge method results always in the lowest epidemic size compared to the other methods. To summarize, if we exclude the case of the Number of neighborhood Communities strategy in the situation where the network has a strong community structure ($ \mu=0.1$), in every other situations the relative difference of the outbreak  is always positive. That  indicates that  all the proposed strategies outperform the alternatives. Let's now turn to more detailed comparisons. First of all, these results clearly demonstrate the superiority of deterministic methods over stochastic methods. Indeed, in any case, the biggest differences are observed with Acquaintance followed by CBF and BHD. In fact, their rank is correlated with the level of information that they possess on the network topology. In fact, Acquaintance is totally agnostic about the network topology, CBF targets the bridges between the communities while BHD targets both bridges and hubs. Even though CBF and BHD are community-based methods, they use the information only at the level of randomly chosen nodes, from where their low performances. The compared effectiveness of the three alternatives deterministic strategies depends on the strength of the community structure. For strong community structure ($ \mu=0.1$), Degree and Comm strategy are very close while Betweenness centrality is slightly more performing whatever the value of fraction of immunized nodes. For medium community structure strength ($ \mu=0.4$), results are more mixed, even if betweenness centrality is still a little bit more efficient. For weak community structure ($ \mu=0.7$), the three strategies are well separated. Degree ranks first, followed by Betweenness and Comm strategy in terms of efficiency.

\subsection{Real-world networks}
\begin{figure*}[t!]
\begin{center}
\includegraphics[width=15cm,height=14cm]{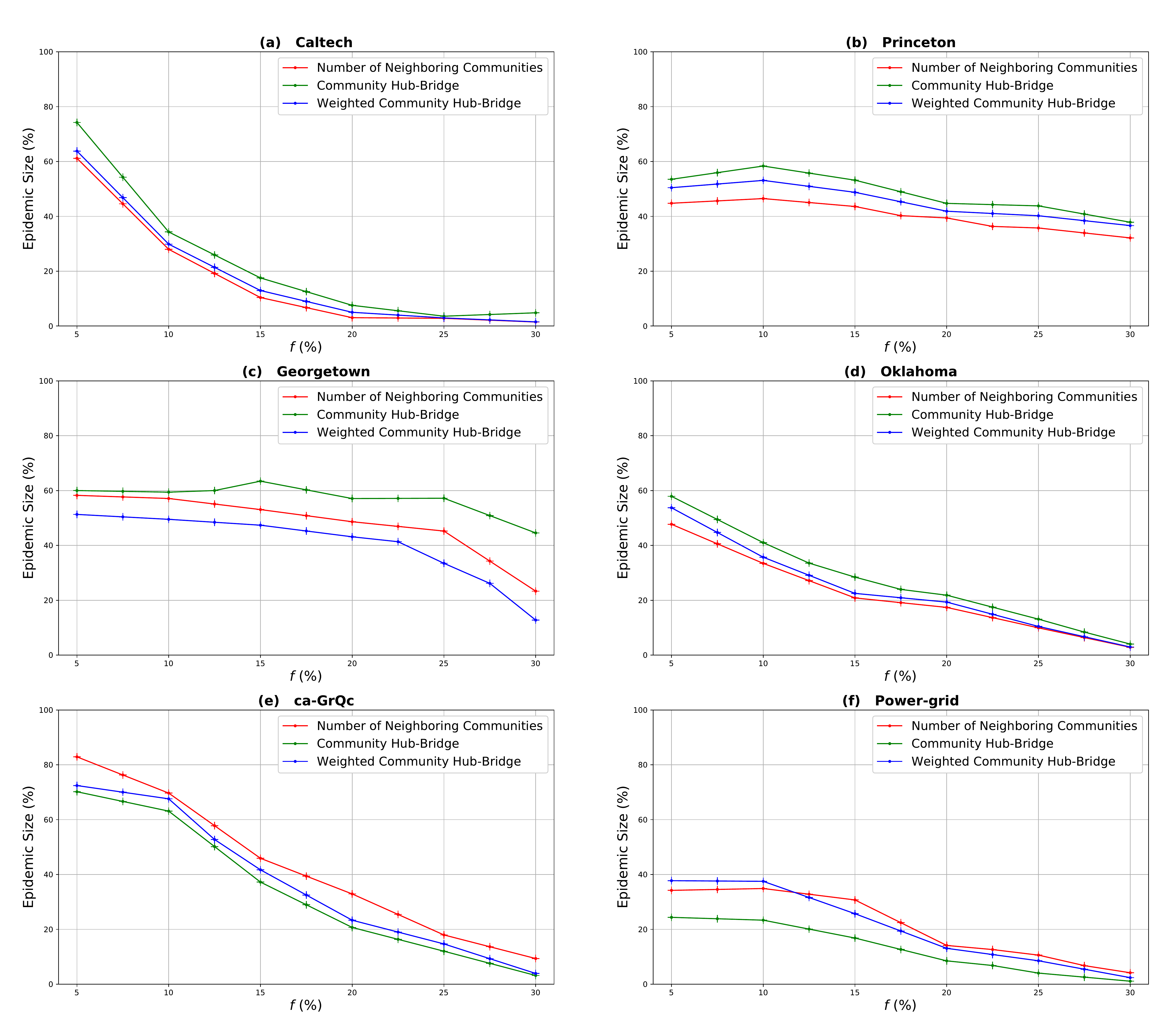}
\caption{The epidemic size of the proposed community based methods performed on six real networks of different types namely on facebook network of four universities (a) Caltech (b) Princeton (c) Georgetown (d) Oklahoma, and on (e) Collaboration network (f) Power network. Final values are obtained by running 600 independent simulations per network, immunization coverage and immunization method.}
\label{r4}
\end{center}
\end{figure*} 
\begin{figure*}[t!]
\begin{center}
\includegraphics[width=15cm,height=20cm]{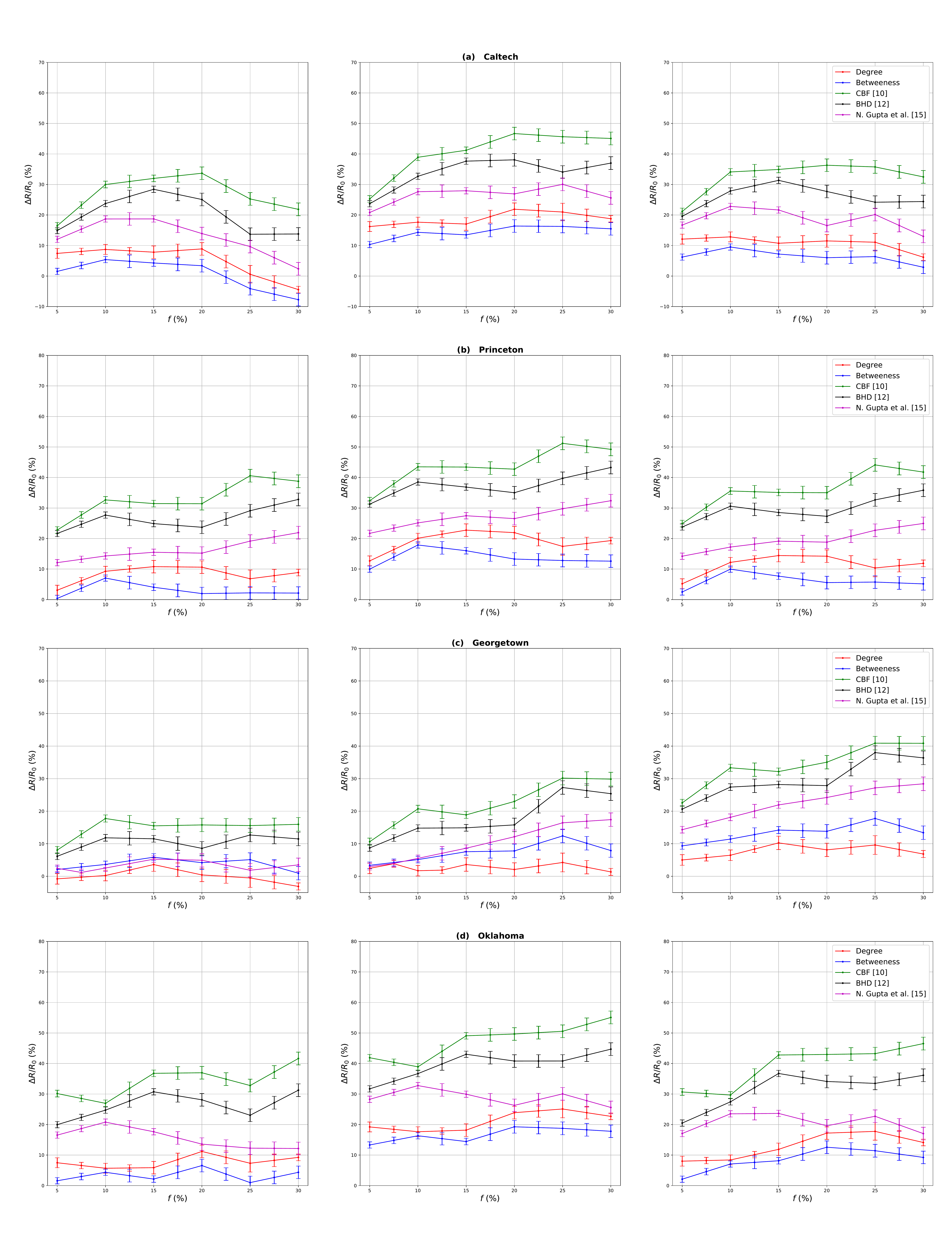}
\label{r51}
\end{center}
\end{figure*}
\begin{figure*}[t!]
\begin{center}
\includegraphics[width=15cm,height=10cm]{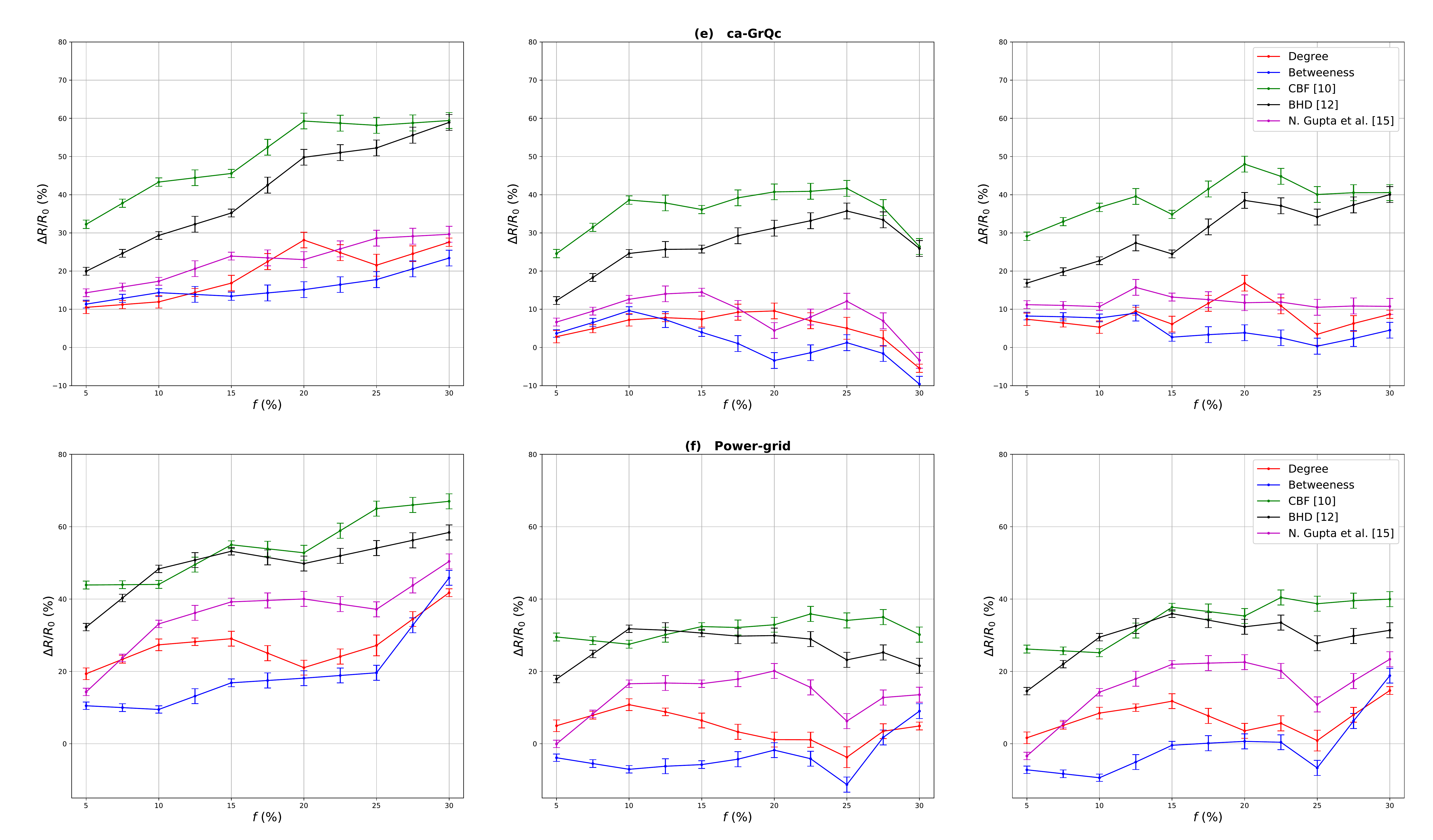}
\caption{The relative difference of the outbreak size $\Delta R/R_{0}$ as a function of the fraction of immunized nodes $f$. The left panels show the difference between Community Hub-Bridge method and the alternative methods. The middle and the right panels show respectively the difference between the Number of Neighboring Communities method, the Weighted Community Hub-Bridge and the alternative methods. We note that a positive value of $\Delta R/R_{0}$ means a higher performance of the proposed method. Simulations are performed on different types of real-world networks. Final values are obtained by running 600 independent simulations per network, immunization coverage and immunization method.}
\label{r5}
\end{center}
\end{figure*}

As our goal is to cover a wide range of situations, real-world data come from different domains (social, technical and collaboration networks). In order to link the results of this set of experiments with those performed on synthetic data, we estimate the mixing proportion parameter of the uncovered community structure by the Louvain algorithm. Indeed, experiments performed with synthetic networks have shown that the community strength is a major parameter in order to explain the efficiency of the proposed immunization strategies. Estimated values reported in \autoref{t3} show that the networks cover a wide range of community strength. 

 \begin{table}
   \centering
    \caption{The estimated mixing parameter $\mu$ of the real-world networks.}
        \label{t3}	
      \begin{tabular}{lcccccc}
      \hline     
	Network & Power-grid & ca-GrQc & Princeton & Oklahoma & Caltech & Georgetown\\
	\hline
	$\mu$ & 0.034 & 0.095 & 0.354 & 0.441 & 0.448 & 0.522\\
	\hline
      \end{tabular}%
\end{table}

\subsubsection{Spreading efficiency of the proposed methods}
\autoref{r4} shows the epidemic size as a function of the fraction of immunized nodes obtained after the SIR simulations for the proposed immunization methods.  These results corroborate the conclusions we made with the synthetic networks. It can be observed on  \autoref{r4} (e) and (f) that in networks with strong community structure, Community Hub-Bridge is the most efficient immunization method.  Indeed, the estimated mixing parameter value  $\mu$ is equal to 0.03 and 0.09 respectively for the power-grid network and  the collaboration network. Communities are very well separated, and the Community Hub-Bridge method targets nodes with a good balance of intra-community and inter-community links. That is where its superiority lies.
 
In networks with average community structure strength shown in \autoref{r4} (a), (b) and (d), the Number of Neighboring Communities outperforms the other proposed methods. It targets the bridges connected to multiple communities which facilitates the spread of epidemics throughout the whole network. Therefore, it is the most efficient method in Caltech, Princeton and Oklahoma networks.

The Weighted Community Hub-Bridge is the most efficient method for the Georgetown network (where $\mu=0.522$) as reported in \autoref{r4} (c). This method depends on the fraction of the inter-community links for each community within the network, which allows us to give the appropriate weighting to favor either the inter-community or the intra-community influence. This is the reason why it outperforms the other proposed methods in the Georgetown network which does not have a strong community structure.
Finally, these results confirm the paramount influence of the mixing proportion parameter in order to choose the most appropriate strategy in a given situation. Whatever the origin of the network, what matters the most is the strength of the community structure.

\subsubsection{Comparison with the alternative methods}

The relative difference of the outbreak size between deterministic as well as stochastic strategies and the proposed strategies is reported in \autoref{r5}. Similarly than with synthetic networks, the stochastic strategies (CBF and BHD) perform poorly as compared to deterministic ones.  Indeed, it appears clearly that these two types of methods are well separated. The results of the comparative evaluation of the deterministic strategies are quite consistent with what might have been expected. The proposed strategies are globally more efficient than their competitors. This is all the more true when they are used appropriately.

The left panels of \autoref{r5} show the comparison between the Community Hub-Bridge and the alternative methods. It outperforms the deterministic methods in networks with strong community structure (Power  Grid and ca-GrQc) with a minimal gain of around 10 \% over the best alternative (Betweenness). Its benefits reduces  when the strength of the community structure gets looser. It is still above Betweenness for Princeton and Oklahoma networks, despite their medium range mixing proportion. However, when the community structure becomes weaker (Caltech and Georgetown), it is less performing when the fraction of immunized nodes is greater than 20 \%.

The middle panels of \autoref{r5} show the comparison between the Number of Neighboring Communities and its alternative. It shows its best performances for networks with medium mixing proportion values (Princeton, Oklahoma, Caltech) with gains above 10 \% as compared to the most performing alternative (Betweenness). It is still performing better than the best alternative for Georgetown (degree strategy), but with gains of less than 10 \%. However, it performs in some cases worse than Degree and Betweenness in networks with strong community structure (Power-grid and the collaboration networks).

The right panels of \autoref{r5} show the comparison between the Weighted Community Hub-Bridge and the alternative strategies. As expected, it outperforms its competitors  in networks with average and high community structure strength. However, it  can be worse than betweenness for networks with a strong community structure.

To summarize, these experiments reveal that the proposed algorithms are very effective in identifying the influential nodes to be selected for immunization. When they are used on the appropriate networks in terms of community strength, they outperform the available strategies, simply by using relevant information about the community structure.
 
\section{Conclusion}
The adoption of an appropriate immunization strategy aroused much interest among researchers to control any threat of 
infectious diseases spreading. Despite the presence of the community structure in all the social networks, this property has 
been mostly ignored by the existing immunization strategies. In this paper, three community based strategies are proposed. They engage more topological information related to networks with non-overlapping community structure. The proposed strategies are evaluated in different synthetic and real networks. To verify their effectiveness, SIR epidemic model is employed.
First of all, results show that the performance of stochastic strategies is far from what can be expected from deterministic strategies. Indeed, as they do not have access to the whole network structure, it is not easy to exploit their properties. 

Extensive investigation shows also that generally, the proposed immunization strategies have a smaller epidemic size compared to the most influential deterministic immunization strategies (Betweenness and Degree) and the Comm strategy designed for networks with non-overlapping community structure. The Community Hub-Bridge method is particularly suited to networks with strong community structure. The Number of Neighboring communities shows its best with medium strength community structure while Weighted Community Hub-Bridge is more efficient in networks with weak community structure. Additionally, the community size range plays also an important role in the diffusion process. Immunization strategies are more efficient when the community size is small.  One of the main benefit  of this work is to show that significant gains can be achieved by making a better use of the knowledge on the community structure organization. It can be extended in multiple directions. First of all, these measures can surely be  improved by using finer weights in order to make them more robust to the community structure variation. Now that the impact of the community structure strength has been clearly identified, stochastic versions of the proposed strategies need to be designed. Finally, extension to non-overlapping community structures can be considered. 


\label{sec:references}

\begin{thebibliography}{00}

%
%

\bibitem{com1} Porter, Mason A., Jukka-Pekka Onnela, and Peter J. Mucha. "Communities in networks." Notices of the AMS 56.9 (2009): 1082-1097.

\bibitem{com2} Ferrara, Emilio. "Community structure discovery in facebook." International Journal of Social Network Mining 1.1 (2012): 67-90.

\bibitem{jebabli1} Jebabli, Malek, et al. "User and group networks on YouTube: A comparative analysis." Computer Systems and Applications (AICCSA), 2015 IEEE/ACS 12th International Conference of. IEEE, 2015.

\bibitem{jebabli2} Jebabli, Malek, et al. "Overlapping community structure in co-authorship networks: A case study." u-and e-Service, Science and Technology (UNESST), 2014 7th International Conference on. IEEE, 2014.

\bibitem{com4} Liu, Zonghua, and Bambi Hu. "Epidemic spreading in community networks." EPL (Europhysics Letters) 72.2 (2005): 315. 

\bibitem{gupta1} Gupta, Naveen, Anurag Singh, and Hocine Cherifi. "Community-based immunization strategies for epidemic control." Communication Systems and Networks (COMSNETS), 2015 7th International Conference on. IEEE, 2015.

\bibitem{chakraborty} Chakraborty, Debayan, Anurag Singh, and Hocine Cherifi. "Immunization Strategies Based on the Overlapping Nodes in Networks with Community Structure." International Conference on Computational Social Networks. Springer, Cham, 2016.

\bibitem{palla} Palla, Gergely, et al. "Uncovering the overlapping community structure of complex networks in nature and society." nature 435.7043 (2005): 814.

\bibitem{fortunato} Fortunato, Santo, and Darko Hric. "Community detection in networks: A user guide." Physics Reports 659 (2016): 1-44.

\bibitem{cbf} Salathé, Marcel, and James H. Jones. "Dynamics and control of diseases in networks with community structure." PLoS computational biology 6.4 (2010): e1000736.

\bibitem{acq} Cohen, Reuven, Shlomo Havlin, and Daniel Ben-Avraham. "Efficient immunization strategies for computer networks and populations." Physical review letters 91.24 (2003): 247901.

\bibitem{bhd} Gong, Kai, et al. "An efficient immunization strategy for community networks." PloS one 8.12 (2013): e83489.

\bibitem{rwos} Taghavian, Fatemeh, Mostafa Salehi, and Mehdi Teimouri. "A local immunization strategy for networks with overlapping community structure." Physica A: Statistical Mechanics and its Applications 467 (2017): 148-156.

\bibitem{centralities} Lü, Linyuan, et al. "Vital nodes identification in complex networks." Physics Reports 650 (2016): 1-63.

\bibitem{comm} Gupta, Naveen, Anurag Singh, and Hocine Cherifi. "Centrality measures for networks with community structure." Physica A: Statistical Mechanics and its Applications 452 (2016): 46-59.

\bibitem{membership} Hébert-Dufresne, Laurent, et al. "Global efficiency of local immunization on complex networks." Scientific reports 3 (2013): 2171.

\bibitem{manish} Kumar, Manish, Anurag Singh, and Hocine Cherifi. "An Efficient Immunization Strategy Using Overlapping Nodes and Its Neighborhoods." Companion of the The Web Conference 2018 on The Web Conference 2018. International World Wide Web Conferences Steering Committee, 2018. 

\bibitem{bet1} NEWMAN, Mark EJ. A measure of betweenness centrality based on random walks. Social networks, 2005, vol. 27, no 1, p. 39-54.

\bibitem{bet2} Brandes, Ulrik. "A faster algorithm for betweenness centrality." Journal of mathematical sociology 25.2 (2001): 163-177.

\bibitem{lfr} Lancichinetti, Andrea, Santo Fortunato, and Filippo Radicchi. "Benchmark graphs for testing community detection algorithms." Physical review E 78.4 (2008): 046110.

\bibitem{orman3} Orman, Günce Keziban, Vincent Labatut, and Hocine Cherifi. "Towards realistic artificial benchmark for community detection algorithms evaluation." International Journal of Web Based Communities 9.3 (2013): 349-370.

\bibitem{barabasi} Albert, Réka, and Albert-László Barabási. "Statistical mechanics of complex networks." Reviews of modern physics 74.1 (2002): 47.

\bibitem{boccaletti} Boccaletti, Stefano, et al. "Complex networks: Structure and dynamics." Physics reports 424.4-5 (2006): 175-308.

\bibitem{newman} Newman, Mark EJ. "The structure and function of complex networks." SIAM review 45.2 (2003): 167-256.

\bibitem{traud} Traud, Amanda L., Peter J. Mucha, and Mason A. Porter. "Social structure of facebook networks." Physica A: Statistical Mechanics and its Applications 391.16 (2012): 4165-4180.

\bibitem{watts} Watts, Duncan J., and Steven H. Strogatz. "Collective dynamics of ‘small-world’networks." nature 393.6684 (1998): 440.

\bibitem{leskovec} Leskovec, Jure, Jon Kleinberg, and Christos Faloutsos. "Graph evolution: Densification and shrinking diameters." ACM Transactions on Knowledge Discovery from Data (TKDD) 1.1 (2007): 2.

\bibitem{orman2} Orman, Günce Keziban, Vincent Labatut, and Hocine Cherifi. "On accuracy of community structure discovery algorithms." arXiv preprint arXiv:1112.4134 (2011).

\bibitem{orman1} Orman, Günce Keziban, Vincent Labatut, and Hocine Cherifi. "Comparative evaluation of community detection algorithms: a topological approach." Journal of Statistical Mechanics: Theory and Experiment 2012.08 (2012): P08001.

\bibitem{sir1} Newman, Mark EJ. "Spread of epidemic disease on networks." Physical review E 66.1 (2002): 016128.

\bibitem{sir2} Moreno, Yamir, Romualdo Pastor-Satorras, and Alessandro Vespignani. "Epidemic outbreaks in complex heterogeneous networks." The European Physical Journal B-Condensed Matter and Complex Systems 26.4 (2002): 521-529.

\end{thebibliography}

\end{document}